\input harvmac
\input psfig
\input epsf
\noblackbox
\newcount\figno
 \figno=0
 \def\fig#1#2#3{
 \par\begingroup\parindent=0pt\leftskip=1cm\rightskip=1cm\parindent=0pt
 \baselineskip=11pt
 \global\advance\figno by 1
 \midinsert
 \epsfxsize=#3
 \centerline{\epsfbox{#2}}
 \vskip 12pt
 {\bf Fig.\ \the\figno: } #1\par
 \endinsert\endgroup\par
 }
 \def\figlabel#1{\xdef#1{\the\figno}}
 \def\encadremath#1{\vbox{\hrule\hbox{\vrule\kern8pt\vbox{\kern8pt
 \hbox{$\displaystyle #1$}\kern8pt}
 \kern8pt\vrule}\hrule}}
 %
 %


 \font\cmss=cmss10
 \font\cmsss=cmss10 at 7pt
 \def\rlx{\relax\leavevmode}
 \def\inbar{\vrule height1.5ex width.4pt depth0pt}
 \def\IC{\relax\,\hbox{$\inbar\kern-.3em{\rm C}$}}
 \def\IN{\relax{\rm I\kern-.18em N}}
 \def\IP{\relax{\rm I\kern-.18em P}}
 \def\ZZ{\rlx\leavevmode\ifmmode\mathchoice{\hbox{\cmss Z\kern-.4em Z}}
  {\hbox{\cmss Z\kern-.4em Z}}{\lower.9pt\hbox{\cmsss Z\kern-.36em Z}}
  {\lower1.2pt\hbox{\cmsss Z\kern-.36em Z}}\else{\cmss Z\kern-.4em
  Z}\fi}
 \def\IZ{\relax\ifmmode\mathchoice
 {\hbox{\cmss Z\kern-.4em Z}}{\hbox{\cmss Z\kern-.4em Z}}
 {\lower.9pt\hbox{\cmsss Z\kern-.4em Z}}
 {\lower1.2pt\hbox{\cmsss Z\kern-.4em Z}}\else{\cmss Z\kern-.4em
 Z}\fi}
 \def\IZ{\relax\ifmmode\mathchoice
 {\hbox{\cmss Z\kern-.4em Z}}{\hbox{\cmss Z\kern-.4em Z}}
 {\lower.9pt\hbox{\cmsss Z\kern-.4em Z}}
 {\lower1.2pt\hbox{\cmsss Z\kern-.4em Z}}\else{\cmss Z\kern-.4em Z}\fi}

 \def\narrowplus{\kern -.04truein + \kern -.03truein}
 \def\narrowminus{- \kern -.04truein}
 \def\narrowminussub{\kern -.02truein - \kern -.01truein}

 \def\frac#1#2{{#1\over #2}}

 \def\IZ{\relax\ifmmode\mathchoice
 {\hbox{\cmss Z\kern-.4em Z}}{\hbox{\cmss Z\kern-.4em Z}}
 {\lower.9pt\hbox{\cmsss Z\kern-.4em Z}}
 {\lower1.2pt\hbox{\cmsss Z\kern-.4em Z}}\else{\cmss Z\kern-.4em Z}\fi}
 \def\IB{\relax{\rm I\kern-.18em B}}
 \def\IC{{\relax\hbox{$\inbar\kern-.3em{\rm C}$}}}
 \def\Ic{{\relax\hbox{$\inbar\kern-.22em{\rm c}$}}}
 \def\ID{\relax{\rm I\kern-.18em D}}
 \def\IE{\relax{\rm I\kern-.18em E}}
 \def\IF{\relax{\rm I\kern-.18em F}}
 \def\IG{\relax\hbox{$\inbar\kern-.3em{\rm G}$}}
 \def\IGa{\relax\hbox{${\rm I}\kern-.18em\Gamma$}}
 \def\IH{\relax{\rm I\kern-.18em H}}
 \def\II{\relax{\rm I\kern-.18em I}}
 \def\IK{\relax{\rm I\kern-.18em K}}
 \def\IP{\relax{\rm I\kern-.18em P}}

 \font\cmss=cmss10 \font\cmsss=cmss10 at 7pt
 \def\IR{\relax{\rm I\kern-.18em R}}

 %

 %
 %
 \def\eqnn#1{\xdef #1{(\secsym\the\meqno)}\writedef{#1\leftbracket#1}%
 \global\advance\meqno by1\wrlabeL#1}
 \def\eqna#1{\xdef #1##1{\hbox{$(\secsym\the\meqno##1)$}}
 \writedef{#1\numbersign1\leftbracket#1{\numbersign1}}%
 \global\advance\meqno by1\wrlabeL{#1$\{\}$}}
 \def\eqn#1#2{\xdef #1{(\secsym\the\meqno)}\writedef{#1\leftbracket#1}%
 \global\advance\meqno by1$$#2\eqno#1\eqlabeL#1$$}

 \lref\author{Name}

\lref\Sinha{S.~Sinha and C.~Vafa,``SO and Sp Chern-Simons at large N,''
arXiv:hep-th/0012136.}
\lref\AKV{
M.~Aganagic, A.~Klemm and C.~Vafa,
``Disk instantons, mirror symmetry and the duality web,''
arXiv:hep-th/0105045.
}
\lref\AV{
M.~Aganagic and C.~Vafa,
``Mirror symmetry, D-branes and counting holomorphic discs,''
arXiv:hep-th/0012041.
}
\lref\kachru{
S.~Kachru, S.~Katz, A.~E.~Lawrence and J.~McGreevy,
``Open string instantons and superpotentials,''
Phys.\ Rev.\ D {\bf 62}, 026001 (2000)
[arXiv:hep-th/9912151];

``Mirror symmetry for open strings,''
Phys.\ Rev.\ D {\bf 62} (2000) 126005
[arXiv:hep-th/0006047].}
\lref\ov{
H.~Ooguri and C.~Vafa,
``Knot invariants and topological strings,''
Nucl.\ Phys.\ B {\bf 577}, 419 (2000)
[arXiv:hep-th/9912123].
}

\lref\vafaaug{
C.~Vafa,
``Superstrings and topological strings at large N,''
J.\ Math.\ Phys.\  {\bf 42}, 2798 (2001)
[arXiv:hep-th/0008142].
}
\lref\gv{
R.~Gopakumar and C.~Vafa,
``On the gauge theory/geometry correspondence,''
Adv.\ Theor.\ Math.\ Phys.\  {\bf 3}, 1415 (1999)
[arXiv:hep-th/9811131].
}
\lref\hv{
K.~Hori and C.~Vafa,
``Mirror symmetry,''
arXiv:hep-th/0002222.
}
\lref\hiv{
K.~Hori, A.~Iqbal and C.~Vafa,
``D-branes and mirror symmetry,''
arXiv:hep-th/0005247.
}
\lref\bcov{
M.~Bershadsky, S.~Cecotti, H.~Ooguri and C.~Vafa,
``Kodaira-Spencer theory of gravity and exact results for quantum string amplitudes,''
Commun.\ Math.\ Phys.\  {\bf 165}, 311 (1994)
[arXiv:hep-th/9309140].
}
\lref\GVW{
S.~Gukov, C.~Vafa and E.~Witten,
``CFT's from Calabi-Yau four-folds,''
Nucl.\ Phys.\ B {\bf 584}, 69 (2000)
[Erratum-ibid.\ B {\bf 608}, 477 (2000)]
[arXiv:hep-th/9906070].
}
\lref\Gukov{
S.~Gukov,
``Solitons, superpotentials and calibrations,''
Nucl.\ Phys.\ B {\bf 574}, 169 (2000)
[arXiv:hep-th/9911011].
}
\lref\VT{
T.~R.~Taylor and C.~Vafa,
``RR flux on Calabi-Yau and partial supersymmetry breaking,''
Phys.\ Lett.\ B {\bf 474}, 130 (2000)
[arXiv:hep-th/9912152].
}
\lref\Mayr{
P.~Mayr,
``On supersymmetry breaking in string theory and its realization in brane  worlds,''
Nucl.\ Phys.\ B {\bf 593}, 99 (2001)
[arXiv:hep-th/0003198].
}
\lref\AS{B. Acharya and B. Spence, ``Flux, Supersymmetry and M theory on 7-manifolds,''
[arXiv:hep-th/0007213].}
\lref\AGD{B. Acharya, X. De La Ossa and S. Gukov,``G-flux, Supersymmetry and
Spin(7) manifolds,'' [arXiv:hep-th/0201227]}
\lref\new{E. Witten, ``New Issues in Manifolds of SU(3) holonomy,'' Nucl.\ Phys.\ B {\bf 268},
(1986) 79.}
\lref\pma{P.~Mayr,
``N = 1 mirror symmetry and open/closed string duality,''
arXiv:hep-th/0108229.}
\lref\govin{S. Govindarajan, T. Jayaraman and T. Sarkar, ``Disc Instantons in Linear Sigma Models,''
[arXiv:hep-th/0108234]}
\lref\lermay{W. Lerche and P. Mayr,``On N=1 Mirror Symmetry for Open Type II strings,'' [arXiv:hep-th
/0111113]}
\lref\Wtoro{E.~Witten,
``Toroidal compactification without vector structure,''
JHEP {\bf 9802} (1998) 006
[arXiv:hep-th/9712028].}
\lref\Blum{
J.~D.~Blum,
``Calculation of nonperturbative terms in open string models,''
arXiv:hep-th/0112039.}
\lref\Iqbal{
A.~Iqbal and A.~K.~Kashani-Poor,
``Discrete symmetries of the superpotential and calculation of disk  invariants,''
arXiv:hep-th/0109214.
}
\Title
 {\vbox{
 \baselineskip12pt
 \hbox{HUTP-01/A073}
 \hbox{RUNHETC-2002-06}
 \hbox{hep-th/0202208}\hbox{}\hbox{}
}}
 {\vbox{
 \centerline{Orientifolds, Mirror Symmetry}
 \vglue .5cm
 \centerline{and}
 \vglue .5cm
 \centerline{Superpotentials}
 }}
 \centerline{ Bobby ${\rm Acharya}^1$, Mina ${\rm Aganagic}^2$, Kentaro
 ${\rm Hori}^2$ and Cumrun ${\rm Vafa}^2$}
 \bigskip\centerline{${}^1$Department of Physics, Rutgers University}
 \centerline{Piscataway, NJ 08854, USA}
 \bigskip\centerline{${}^2$Jefferson Physical Laboratory}
 \centerline{Harvard University, Cambridge, MA 02138, USA}
 \smallskip
 \vskip .3in \centerline{\bf Abstract}
{We consider orientifolds of Calabi-Yau 3-folds in the
context of Type IIA and Type IIB superstrings.  We show how mirror symmetry
can be used to sum up worldsheet instanton contributions
to the superpotential for Type IIA
superstrings.  The relevant worldsheets have the topology
of the disc and ${\bf RP^2}$.}
 \smallskip
\Date{}
\newsec{Introduction}

Mirror symmetry has been proven effective in the computation
of superpotentials in four dimensional
${\cal N}=1$ supersymmetric
theories \refs{\AV,\AKV,\pma,\govin,\Iqbal,\lermay,\Blum}. In this
paper we show how mirror symmetry can be used in the context
of orientifolds of Calabi-Yau backgrounds to lead to computable superpotentials
which receive corrections from worldsheet instantons.  The
worldsheet instantons have the topology of ${\bf RP^2}$ and, if there are
D-branes around, the disc ${\bf D^2}$. In principle both contribute
to the superpotential, as we will discuss.  Along the way, we
confirm a prediction of \Sinha\ (based on a large $N$ Chern-Simons
conjecture) for the superpotential
arising  when we consider Type IIA superstrings
propagating on an orientifold of the resolution of the conifold.

The organization of this paper is as follows:  In section 2 we describe some
general features of supersymmetric orientifolds of Type IIA and IIB string
theories on Calabi-Yau manifolds.
We describe the spacetime superpotential in the case of orientifolds of IIB.
In section 3 we present examples of Type IIA orientifolds of non-compact
Calabi-Yau threefolds, as well as their Type IIB mirrors.
In section 4 we present further examples of non-compact CY orientifolds.

\newsec{Superstrings and Calabi-Yau Orientifolds}

Compactification of Type IIA or IIB superstring theory  on a Calabi-Yau
threefold $X$ gives rise to an ${\cal N}=2$ supersymmetric theory
in $d=4$.  We can also consider orientifolding this theory, by
combining an involution symmetry $I$ on the Calabi-Yau with
orientation reversal $\Omega$ on the worldsheet, i.e. we
consider gauging the symmetry generated by $I\Omega$.
Due to their respective orientations, this orientifold is only well defined
if $I$ preserves the orientation of $X$ in the case of the IIB string and reverses it in the
case of IIA. Additionally, if $I$ obeys certain further properties, the orientifold theory
will be an ${\cal N}=1$
supersymmetric theory in $d=4$:
For Type IIA superstrings
the involution $I$ has to exchange the holomorphic 3-form
$\omega$, up to a phase, with the anti-holomorphic 3-form
$\overline \omega$.  This is  because the left-moving
space-time supercharge
corresponds to the holomorphic 3-form and the right-moving space-time
supercharge corresponds to the anti-holomorphic 3-form.  This means
that $I$ should be an anti-holomorphic involution in the context
of Type IIA superstrings.  For type IIB superstrings both
the left- and right-moving supercharges correspond to the holomorphic
3-form on Calabi-Yau.  Thus the involution should send $\omega
\rightarrow \pm \omega$.  This is a holomorphic involution.

If $I$  has fixed points $\Sigma$ in $X$, we find orientifold
planes of topology $\Sigma{\times}{\bf R^4}$.
For Type IIA superstrings, the fixed point of the anti-holomorphic
involution will be real codimension three.
In other words, it is an O6 plane.
To see this, note that in local coordinates on a point
on orientifold plane an anti-holomorphic involution will have three $-1$ eigenvalues
and three $+1$ eigenvalues.  For Type IIB, the orientifold plane could
be O3, O5, O7 or O9, depending on the number of eigenvalues of the holomorphic
involution on the tangent space of any point on the orientifold plane.
Note that to have an O5 or O9 the involution takes
$\omega \rightarrow \omega$, and for O3, O7 it takes $\omega \rightarrow
-\omega$.  Note that we can have one or the other of these situations if we wish
to preserve supersymmetry, but not both.

In compact Calabi-Yau threefolds, if we end up
with orientifold planes, we need to cancel their D-brane charge
by including appropriate space-time filling, Calabi-Yau wrapped D-branes.
In the language of the `parent' theory before orientifolding,
this corresponds to including wrapped D-brane
configurations which are invariant under the involution $I\Omega$.
In the non-compact case, we do not need to cancel the D-brane charge
(except for the D9 brane charge) as the flux can go off to infinity.
Nevertheless we can choose to include D-branes in the background.

Note that since Type IIA and IIB superstrings on mirror Calabi-Yau pairs
are equivalent, it follows that the orientifold operation of IIA
is equivalent to an orientifold operation for IIB on the mirror Calabi-Yau.
In this case the O6-plane gets mapped to O3- and O7-planes
${\it or}$ O5- and O9-planes.  Similarly D6-branes get mapped to D3- and D7-branes
or D5- and D9-branes depending on which case we are in.

\subsec{Space-time superpotential}
Since we have ${\cal N}=1$ supersymmetry in $d=4$ a
superpotential can in principle be generated by the orientifold operation.
General arguments \refs{\GVW, \Gukov} which relate the tensions of domain walls
to the value of the superpotential can be used to propose a formula for
the superpotential itself in various classes of string compactifications.
These formulae have proven successful in Calabi-Yau compactifications \refs{\GVW,\VT,\Mayr},
compactifications on $G_2$-manifolds, \refs{\AS} and even $Spin(7)$-manifolds \refs{\AGD}.
In the context of orientifolds of Type IIB string theory on Calabi-Yau threefolds, the BPS domain
walls are actually D5-branes wrapped on supersymmetric 3-cycles in X. The fact that such 3-cycles
are in fact calibrated by $\omega$ leads to the following formula for the superpotential,
\eqn\suppoh{W=\int H \wedge \omega}
where $H=dB_{RR}$ denotes the RR 3-form field strength. As a check on this formula, note that
in Type I theory, $H$ also includes contributions from the Chern-Simons 3-forms constructed
out of the $SO(32)$ gauge field $A$ and the spin connection $w$. Thus, $W$ contains a term
which is none other that the holomorphic Chern-Simons functional,
\eqn\csw{\int tr {(A \wedge dA + {2 \over 3} A \wedge A \wedge A ) \wedge \omega}}
This is the known superpotential for Type I and heterotic string theory on a Calabi-Yau threefold
\refs{\new}. In fact this is simply the expected superpotential
from the viewpoint of topological B-model on the 9-brane
\ref\wcss{E. Witten, ``Chern-Simons Gauge Theory As A String Theory,''
hep-th/9207094.}\bcov .  Moreover, the fact that
$H$ is defined with the addition of the Chern-Simons 3-form in the type
I theory,
reflects that fact that an instanton on the 9-brane can be viewed
as a 5-brane, which is the source of $H$-flux.

Since both D5-branes and O5-planes
are sources for $H$, we can think of  $W$ as being ``generated'' by D5-brane charge.
All sources for $H$ contribute.
Let $D_i$ denote
the locations of the D5-brane charges (which includes O5-planes).
Each $D_i$ is a 2-dimensional
subspace of the Calabi-Yau. This is because locally we have that
\eqn\dcont{dH = \sum_i \delta_i }
where the $\delta_i$ are four form currents supported at the
D5 or O5 locations. These are Poincare dual to 2-cycles $D_i$.

Suppose we are considering the compact case.
Then the total 5-brane charge is zero, which implies that $\sum_i [D_i]=0$
where $[D_i]$ denotes the  class in ${H_2}(X, {\bf Z})$ of $D_i$.
This implies that there is a three dimensional
subspace of the Calabi-Yau, $C$, such that its boundary is
$$\partial C=\sum_i D_i$$
Then we can view $C$ as the ``flux tube'' for the D5-brane charges.
Of course $C$ is not unique.  However the difference of any two choices
gives a closed 3-cycle in the Calabi-Yau, which corresponds to turning
on an integer $H$-flux in the Calabi-Yau.  So for a given
configuration of $H$-fluxes in the bulk Calabi-Yau, we get a
class of $C$'s, whose difference is homologically trivial.  The fact
that $C$ is a flux tube for D5 brane charge means that
$$\int H\wedge \alpha= \int_C \alpha$$
for any closed 3-form $\alpha$.  Applying this to \suppoh\ we find
\eqn\funda{W=\int H\wedge \omega =\int_C \omega}
where $C$ is a 3-chain with $\partial C=\sum_i D_i$.
This formula was derived in the compact case, but the idea
also applies to the non-compact
case.
In that case we do not
need a net 5 brane charge being zero, as the flux can go to infinity.
Put differently we can assume we have a non-compact $C$ by taking
the boundary at infinity corresponding to ``5-brane charge at infinity''.
Thus the same formula still applies.  More precisely, in that case we
can view \funda\ as defining the superpotential up to an addition
of a constant corresponding to how we fix the boundary condition at
infinity \lref\Wb{E.~Witten,``Branes and the dynamics of {QCD},''
Nucl.\ Phys.\ B {\bf 507}, 658 (1997)
[arXiv:hep-th/9706109].}
\refs{\Wb,\AV,\AKV}.

What does this correspond to at the level of worldsheet computations?
The $D_i$ correspond to charges coming from physical
D5 branes as well as O5 planes
carrying D5 brane charge.  As far as the computations of the
physical D5 branes they correspond to contributions from worldsheet
geometry being a disk  \lref\dougl{
I.~Brunner, M.~R.~Douglas, A.~E.~Lawrence and C.~Romelsberger,
``D-branes on the quintic,''JHEP {\bf 0008}, 015 (2000)
[arXiv:hep-th/9906200].}
\refs{\bcov,\dougl,\kachru,\AV} . As
far as the contribution to the superpotential due to the O5 planes, this
should come from non-orientable worldsheets.  Based on
the requisite R-charge, it is possible to show
that they can only come from ${\bf RP^2}$.

To see this, note that for
worldsheet geometries being an ${\bf S^2}$, in the Berkovits formalism,
computation of F-type terms leads to
 4 fermionic zero modes on the worldsheet which contribute
 to space-time action
$$\int d^4\theta F(t_i)$$
where $F$ is the prepotential of $N=2$ theory in $d=4$ and $t_i$
denote the vector multiplet. Moreover $F$ is the partition
function of topological strings on Calabi-Yau where $t_i$ is identified
with the moduli of Calabi-Yau.  If we consider ${\bf RP^2}$ geometry
instead, the same analysis leads to only two fermionic zero modes
(two being gotten rid of by the orientifold operation).  Thus
we end up with an ${\cal N}=1$ superpotential term \Sinha
$$\int d^2 \theta W(t_i)$$
where $t_i$ denote chiral fields which survive from the
corresponding vector
multiplet under the orientifold operation.  Here $W(t_i)$ corresponds
to the partition function of topological strings on ${\bf RP^2}$.  In this
case, it corresponds to the partition function of topological B-model
on ${\bf RP^2}$.

We can also ask how these worldsheet computations arise in the
context of Type IIA orientifolds.  By mirror symmetry, the worldsheets
will arise in the same way:  Namely there will be contribution
to the superpotential corresponding to disk
amplitudes (with boundary on D6 branes)
and there will also be a contribution to the superpotential
from ${\bf RP^2}$.  The superpotential in the IIA string theory can be
computed by the topological
A-model theory (The B-model computation that is relevant for the IIB
superpotential ends up being equivalent to that described above \AV .).
Recall that the topological A-model,
roughly speaking, counts the number of holomorphic maps from the worldsheet
to the target, weighted by $e^{-A}$ where $A$ is the area of the image.
In the context of non-orientable worldsheet geometries the same
is true when we go to the covering theory.  In other words we consider
the worldsheet geometries with crosscaps to be the ${\bf Z_2}$ quotients of an
 orientable
Riemann surface and we count holomorphic maps from the covering
worldsheet to the target, compatible with the
simultaneous ${\bf Z_2}$ actions on the worlsheet
and the target (in mathematical terminology these are known
as ${\bf Z_2}$ equivariant maps).  In this way, once we know
the mirror geometry, we can use the Type IIB result
given by \funda\ to obtain the non-trivial worldsheet instanton
generated superpotential
in the Type IIA setup.  At the level of disk amplitudes this
was already done in \refs{\AV ,\AKV}.  Here we will be mostly
interested in the contribution from the ${\bf RP^2}$ diagrams
to the superpotential, which correspond, in the Type IIB
setup, to the contribution of the O5 planes to the superpotential.

We now turn to examples.

\newsec{Examples}

In this section we present a number of examples corresponding
to IIA and IIB superstrings on orientifolds
of Calabi-Yau threefolds.  We consider non-compact
models, starting with examples of two distinct orientifolds of the
resolved conifold in the IIA setup.  We also study
its type IIB mirror and use that to compute the contribution
of worldsheet
instantons of the IIA theory with the topology of ${\bf RP^2}$
to the superpotential.  We then generalize the discussion to
some other non-compact examples.

\subsec{Anti-holomorphic orientifolds of $O(-1)+O(-1)\rightarrow \bf{CP^1}$}

We start with an example of local A-model geometry, a Calabi-Yau manifold
$X$ that is an
$O(-1)+ O(-1)$ bundle over ${\bf CP^1}$.

The Calabi-Yau sigma-model can be obtained as the theory on
the Higgs branch of a linear sigma model in two dimensions with $N=(2,2)$
supersymmetry \ref\phases{
E.~Witten,``Phases of N = 2 theories in two dimensions,''
Nucl.\ Phys.\ B {\bf 403}, 159 (1993)
[arXiv:hep-th/9301042].}. The linear sigma model associated
to $X$ has gauge group $G=U(1)$ and two chiral fields
$X_{1,2}$ of charge $+1$
and two fields $X_{3,4}$ of charge ${-1}$.
The Higgs branch is the space of minima of D-term potential:
$$|X_1|^2+|X_2|^2-|X_3|^2-|X_4|^2=r,$$
modulo $G$ action.
As described in \AV\ , $X$ can naturally be viewed
as a $T^3$ fibration over a toric base.
The size of the $\bf{CP^1}$ is set by the
Fayet-Illiopolous parameter $r$. This is complexified
by the theta angle of the gauge theory to give
the complexified Kahler parameter $t=r-i \theta$
on which the A-model amplitudes depend.

Worldsheet orientation reversal
is a symmetry of topological A-model
when accompanied with an anti-holomorphic
involution of the target space.
This is because the A-type twist
correlates the chirality of the fermions and the $U(1)_R$ charge
in such a way that, e.g.,  $\psi_{-}^{i}, \psi_{+}^{\bar i}$
have the same topological charge.
Orientation reversal $\Omega$ exchanges left and right moving
fermions so in the A-model, it must be accompanied with an anti-holomorphic
involution $I$ of the target space.  This is in accord
with the Type IIA superstring interpretation discussed
in the introduction.

As anti-holomorphic involutions we
take the following possible actions on chiral superfields
\eqn\inv{I^{\pm}\quad:(X_1,X_2,X_3,X_4)\rightarrow
(\bar{X}_2,\pm \bar{X}_1,\bar{X}_4,\pm \bar{X}_3).}
The involution clearly commutes with $G$, and can be extended to
a symmetry of the linear-sigma model Lagrangian for any value of the
complexified Kahler parameter $t$.
If we define $z=X_1/X_2$, coordinatizing the ${\bf CP^1}$, then
these involutions act
$$I^{\pm}: z\rightarrow \pm{1\over \overline z}$$

These are the unique, up to diffeomorphism, involutions of ${\bf CP^1}$.
$I^-$ has no fixed points and the quotient of ${\bf CP^1}$ by $I^-$ is
${\bf RP^2}$. Therefore, $I^-$ has no fixed points on the bundle over
${\bf CP^1}$.
$I^+$ has a circle of fixed points in ${\bf CP^1}$ given by
$|z|=1$. This circle is naturally regarded as a copy of ${\bf RP^1}$ $\subset$
${\bf CP^1}$.
Its fixed points in  the
$O(-1)+ O(-1)$ bundle over ${\bf CP^1}$ are easily seen to be
a bundle over ${\bf RP^1}$. The fibers are all copies of
${\bf R^2}$.  Thus in this case we get an O6-plane $L$ which is
this rank two real bundle over ${\bf RP^1}$. To be completely
explicit, the fixed point set
is given by
\eqn\fps{|X_1|^2 = |X_2|^2, \quad\quad |X_3|^2 = |X_4|^2, \quad\quad
\sum_{i=1}^{4}\theta_i=0.}
and modulo $G$ action and the D-term equation.
 Here $\theta_i$ is the phase of $X_i$, $X_i= |X_i|e^{i
\theta_i}$.
By construction, the fixed point set must be a Lagrangian submanifold of
the Calabi-Yau. In fact it
is in the family of Lagrangians constructed in
\AV . There, a 1-dimensional moduli space of Lagrangians
was found. This family is $I^+$-invariant at just one point - which
is exactly the 3-submanifold $L$ described here.
See fig 1.

There is an alternative description of both $X$ and the orientifold action
in terms of coordinates invariant under complexified gauge group $G_C =
\bf{C^*}$. These are $x_{13}, x_{24},x_{14},x_{23}$,
where $x_{ij} = X_{i}X{_j}$, and they satisfy
one constraint $$x_{13}x_{24}=x_{14}x_{23},$$
which is the equation of the conifold.

Introducing a coordinate $z=X_1/X_2$
$$x_{14} = z x_{24},\quad x_{13} = z x_{23},$$
solves $x_{ij}=0$, and
defines transition functions of $O(-1)+O(-1)\rightarrow {\bf P^1}$,
where $z$ is local coordinate on the ${\bf CP^1}$ and  $x_{13},x_{14}$
coordinates on fibers. The involution acts as
The involution acts as $ I^{\pm}: (x_{14},x_{13}, x_{24},x_{23})
\rightarrow (\pm {\bar x}_{23}, {\bar x}_{24}, {\bar x}_{13},\pm {\bar
x}_{14}),$ or
$$I^{\pm} \quad : (z,x_{13},x_{14}) \rightarrow ( \pm 1/\bar z,\bar
x_{14}/\bar z, \pm \bar x_{13}/\bar z).$$
The involution $I^-$ has no fixed points as noted before.
The fixed point set for $I^+$ is $|z|=1$ and
${\bar z} x_{14}={\bar x}_{13}$.

\subsec{Mirror B-model geometry}

Mirror symmetry exchanges the sign of the left moving $U(1)_R$ charge, so
it maps the anti-holomorphic involution of
the A-model into a holomorphic involution of the B-model.
This is compatible with the target space interpretation
where we expect for Type IIB that the orientifold operation will
involve a holomorphic involution on the Calabi-Yau.

The mirror of $O(-1)+O(-1)\rightarrow {\bf P^1}$ \hv\ is
a Landau-Ginzburg theory in terms of four fields $Y_i$ that are
constrained
by
$$Y_1+Y_2-Y_3-Y_4 = t,$$
and a superpotential
\eqn\supw{W=e^{-Y_1}+e^{-Y_2}+e^{-Y_3}+e^{-Y_4}.}
The fields $Y_i$ are periodic, $Y_i\sim Y_i +2\pi i$. They are
obtained \hv\ by dualization of the phases of
linear sigma model fields $X^{i}$, and there is a relation
\eqn\mm{Re(Y_i) =|X_i|^2.}
Above, $t$ is the complexified Kahler parameter of the A-model
$t= r-i\theta$.

At the level of the topological theory, the mirror can equivalently be
thought of as a theory
of variations of complex structures
of a certain hypersurface $Y$ \hiv\ .
The B-model has a sigma-model description based on
\eqn\mirrb{Y:\quad xz = e^{-u}+e^{u-v-t}+e^{-v}+1.}
Moreover the four terms on the right-hand side above arise
from the four monomials in the superpotential.  More precisely,
equation \mirrb\
arises by writing
$Y_1 = u+\lambda,
Y_2= -u+v +t+\lambda,Y_3 =v+\lambda,Y_4=\lambda$.
Choosing now projective coordinates
gives the right-hand side of \supw.

The anti-holomorphic involution maps $|X_1|^2 \rightarrow |X_2|^2$
and $|X_3|^2 \rightarrow |X_4|^2$, so the mirror map \mm\ implies
that the mirror holomorphic involutions ${\hat I}^{\pm}$ {\it both} act as
$${\hat I}^{\pm}\quad:(Y_1,Y_2,Y_3, Y_4)
\rightarrow (Y_2 + i \pi ,Y_1 + i \pi ,Y_4 + i \pi ,Y_3 + i\pi).$$
More precisely, \mm\ and holomorphy fix the action up to additions of
$i\pi$ . But  ${\hat I}^{\pm}\Omega$ must be a symmetry of the theory,
and for this
the superpotential $W$ must be odd under ${\hat I}^{\pm}$.
This is because $W$ enters the Lagrangian of the Landau-Ginzburg theory
as $\int d\theta^{+}d\theta^{-} W$, and  since
$\Omega$ acts by exchanging superspace fermionic coordinates
$\theta^+$, and  $\theta^-$, ${\hat I}^{\pm}$ must take $W\rightarrow -W$.
{}From this we learn that the mirror involutions {\it both} act
as
$$(x,z,e^{-u},e^{v})\rightarrow (-x, -z e^v, e^{u-t},e^{-v}).$$

The action on $x,z$ follows from the projectivization
of the superpotential--actually from this it follows that $xz\rightarrow
xze^{v}$ and given the periodicity of $v$ we have
to take either $x$ or $z$ to go to itself with an extra $e^v$ factor.
The choice of sign on $x,z$ is not apriori determined.  The choice
made above turns out to be the correct one, as we shall see.
The action on the other variables
follows from the relation of the $u,v$ to the $Y_i$ and how the orientifold
action acts on the $Y_i$.

The fixed point
set (i.e. the IIB orientifold plane) is given by  solutions to
$$x= - x, \quad z(1+e^v) =0,\quad  2u-t=0 , \quad 2v=0$$
which has only the two components
$x=0, v=i\pi, u=t/2 , t/2+i\pi,$
generically and these are complex curves parameterized by $z$.
There is another solution when the manifold $Y$ becomes singular, at
$t=0$, which is given by
$x=0, z=0, v=i\pi, u =0$. This corresponds to a new orientifold fixed plane
at the conifold singularity. We
will not need it for purposes of this paper.

To summarize, we have two orientifold five-planes that are
two holomorphic curves parameterized by
$z$, and located at
$$x=0,\quad v=i\pi, \quad u=-t/2 \;\;and\;\; -t/2+i\pi.$$
This is consistent with the results of \AV\ which showed
that the D6 branes wrapping $L$ map to D5 branes wrapping
holomorphic curve given by $x=0$, and a choice of a point on the
Riemann surface $0= e^{-u}+e^{u-v-t}+e^{-v}+1$. This, in retrospect,
justifies the choice of sign in the orientifold action on
the $x,z$ we have made above.

\subsec{Proposal for Mirrors}

Note that two distinct orientifold actions on the A-model geometry
are mirror to a single orientifold action in the B-model.
However, to fully specify the theories with orientifolds we must
decide on the signs of the corresponding cross-cap states.
If there is an orientifold plane, the sign determines its RR-charge.
In the Type IIA side, for each of $I^+\Omega$ and $I^-\Omega$
there are two possibilities.
For $I^+\Omega$, in one theory $L$ is an $O6^-$-plane,
while it is $O6^+$ in the other.
For $I^-\Omega$, there is no orientifold plane
but still there are two possibilities.
On the other hand,
in the Type IIB side there are two O5-planes as noted above.
Accordingly, there are four physically distinct
possibilities corresponding to
$++$,$--$,$-+$ and $+-$ charges.

We propose that the IIB orientifold with
$+-$ and $-+$ charges are mirror to the two orientifolds
of Type IIA by $I^-\Omega$ and that the
$--$ and $++$ theories are respectively mirror to IIA with
$O6^-$ and $O6^+$. This is the only assignment of mirrors consistent
with the total RR charges being the same in both theories.
The total D5 charge of $--$ ($++$) is $-2$ ($+2$)
which is the total D6-brane charge of $O6^{-}$ ($O6^+$).
On the other hand, $+-$ and $-+$ theories
both have zero total fivebrane charge which is
mirror to the statement that the orientifold of IIA by $I^-\Omega$ has zero
sixbrane charge.
Note also that $I^-$ acts on the circle $|z|=1$ as the shift by half
period. It is known that the orientifold
of $S^1$ by half-period shift is sent by T-duality to
the orientifold of the dual $S^1$ with two fixed planes
of the opposite signs \Wtoro.
Our proposal is consistent with this fact.


\subsec{Computation of space-time superpotential}
As mentioned in section 2, the computation of the space-time superpotential
for orientifolds of Calabi-Yau threefolds will involve the
computation of topological string amplitudes with the topology of
${\bf RP^2}$.  Moreover, if we add D-branes to this setup,
there will be additional contributions to the superpotential
which can be computed by evaluating
the topological disk amplitudes as in \AV .  Here we will
concentrate mainly on the contribution from ${\bf RP^2}$ to the superpotential
as the disk amplitudes have been already well studied
in the context of mirror symmetry \refs{\AV , \AKV}.

In the context of orientifolds of the A-model \Sinha, the relevant maps
for the ${\bf RP^2}$ worldsheet
are holomorphic maps from ${\bf CP^1}$ to Calabi-Yau
which are ${\bf Z_2}$-equivariant.  Let $z$ denote the coordinates of the
${\bf P^1}$ in the target $O(-1)+O(-1)$ geometry and
$w$ denote the coordinate of the ${\bf P^1}$ on the
worldsheet.  We are thus looking for maps
$z(w)$ such that $$z(-1/ {\overline w})=\pm 1/{\overline z}$$ where
$\pm$ depends on which orientifold $I^{\pm}$ we are considering.
Examples of such maps include $z=w^{2n}$ for $I^+$ and $z=w^{2n+1}$
for $I^-$.  It is not difficult to show that quite
generally the parity of the degree of the map
is correlated with $\pm$ choice in $I^{\pm}$ as the above
representative maps suggest.  Therefore for the ${\bf RP^2}$ amplitude
one expects an infinite sum over odd or even degrees of maps
depending on whether one is considering $I^-$ or $I^+$.  We will
now see that this follows from the mirror description we have
obtained.

Mirror symmetry implies the equivalence between topological A-model amplitudes
on the Calabi-Yau and topological B-model amplitudes on the mirror.
In particular, as noted in section 2, the computation of the superpotential
corresponds to computing the integrals of the holomorphic
3-form \funda\ on 3-chains with boundaries given by D-brane
charges.  In the local models under consideration it was shown in
\AV\ that this computation reduces to certain definite integrals (of  the
 Abel-Jacobi map) on the Riemman surface
$$0 = e^{-u}+e^{-v}+e^{u-v-t}+1.$$
The integral is of the form $\int_\gamma \lambda$
where $\lambda$ is a particular one form $vdu$ and $\partial \gamma=$
points where the D5 brane charges are localized on the Riemann
surface, including the sign and multiplicity of the D5 brane charge.
This charge can arise from physical D5 branes or O5 planes
carrying D5 brane charge. In other words, the integral over the three chain $C$
reduces to an integral over the 1-chain $\gamma$ on the Riemann surface.

As discussed above, the mirror of $I^+$
orientifold are two $O5$ planes of the same charge, and the mirror of
$I^-$ orientifold are two $O5$ planes of opposite charge.
The contribution of the orientifold $++$ and $+-$ five planes
to the superpotential is thus
$$W =\int_{u^*}^{u=t/2} v(u) du\pm\int_{u^*}^{u=t/2 + i \pi} v(u) du.$$
where we have put $\pm$ on the second term because the
choice of the $D5$ brane charge on the second orientifold
plane depends on whether we are considering $I^{\pm}$.
 Note that
each O5 plane carries $\pm 1$ unit of D5 brane charge.
The above formula gives $W$ up to a
choice of an arbitrary base point $u^*$ on the Riemman surface.
The need to pick $u^*$ is due to non-compactness of the D-brane,
and corresponds to the boundary condition at infinity \AV \ .
Explicitly,
\eqn\suppo{\eqalign{{W} & =  \;\int^{t/2}_{u^*}\log\frac{1-
e^{u-t}}{1-e^{-u}}du \pm  \int^{t/2+ i \pi}_{u^*}
\log\frac{1- e^{u-t}}{1-e^{-u}}du\cr
&= - 2  \sum_{n=1}\{1\pm (-1)^n\} \frac{e^{- n t/ 2}}{n^2}}.}
The term in the sum weighted by $e^{-n t/2}$
in the language of topological A-model comes from
a map that is an $n$-fold cover of ${\bf RP^2}$, as
the action of the instanton wrapping ${\bf CP^1}/{\bf Z_2}$ once is one half the
size of the ${\bf CP^1}$ in the covering space.
Moreover we see that for $I^+$ we get contributions only
from even degree whereas for $I^-$ we get contributions
only from odd degrees, as was expected.

In \Sinha\ it is predicted that the
A-model amplitudes coming from unorientable worldsheets have integrality
properties analogous to A-model amplitudes on Riemman surfaces
with boundaries. The two factors above then correspond
to two BPS bound states of D2 branes wrapping the ${\bf{CP^1}}$
in the covering space and ``ending'' on the orientifold, in the sense that
due to
orientifolding, they propagate
only in
two dimensions. Note that their contributions to the partition function
are exchanged, up to an overall sign, by $t \rightarrow t + 2\pi i$.
This has the interpretation \Sinha\ that the two states differ by
$1/2$ unit of $D0$ brane charge. This is so as the sum over
$n$ corresponds to existence of a bound state with $n$
D0 branes for every primitive BPS state. Shifting $t$ by $2\pi$
corresponds to shifting the $B$ field through the ${\bf CP^1}$
by an amount corresponding to $1$ unit of $D0$ brane.

\bigskip
\centerline{\epsfxsize 4.0truein\epsfbox{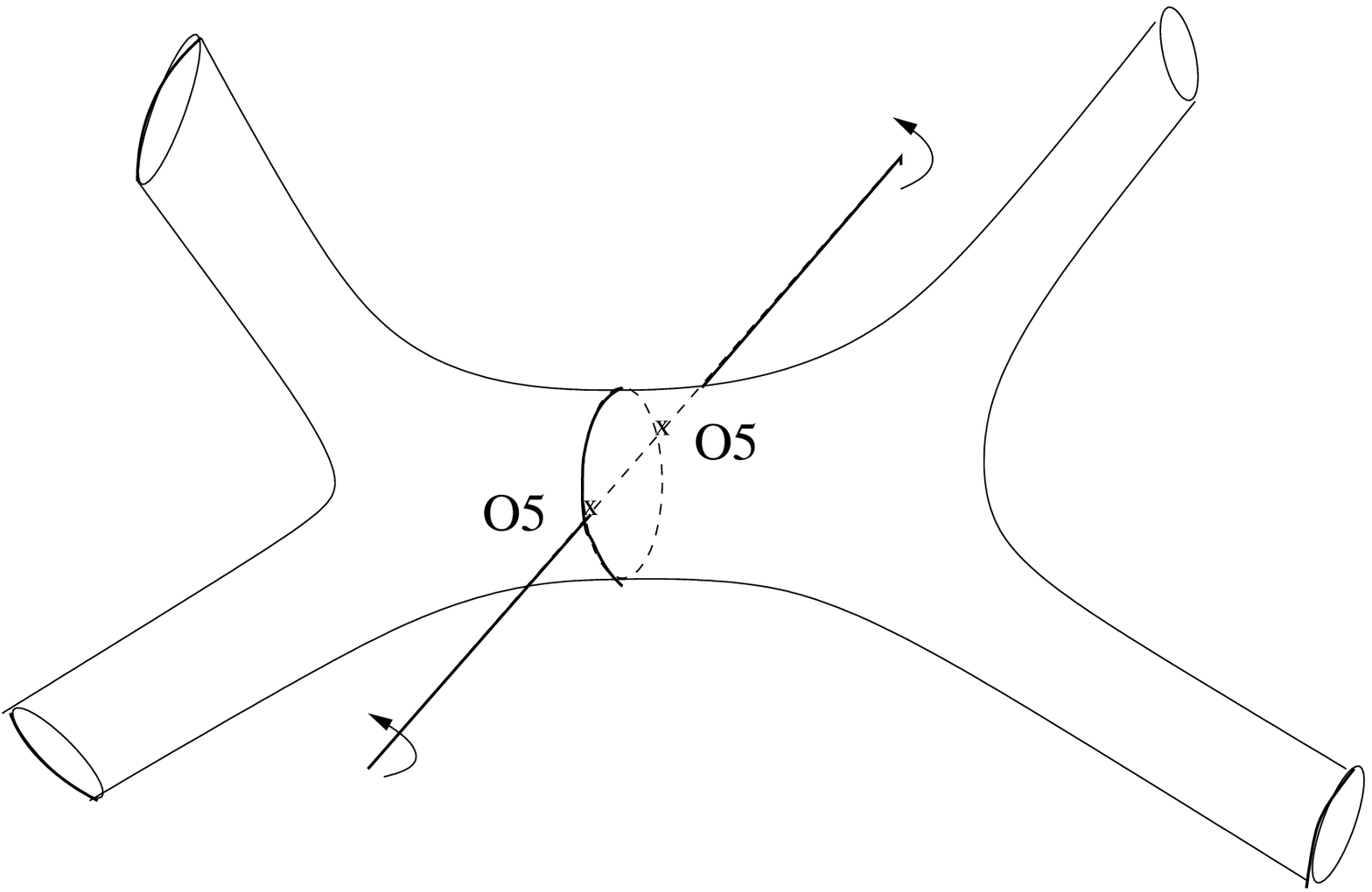}}
\rightskip 2pc \noindent{\ninepoint\sl
\baselineskip=8pt {\bf
Fig.1}: The mirror orientifold action has two fixed points on the
Riemman surface $e^{-u}+e^{-v}+e^{-t+u-v}+1 =0$, at $u=t/2$ and
$u=t/2+i\pi$.
This corresponds to orientifold five-planes in the full geometry.
The possibilities for charge assignments -- equal or opposite charges --
correspond
to different orientifold actions in the A-model.}
\bigskip

We can also add D6 branes to the A-model in which case the
disk amplitude will also be non-vanishing. We can compute this as in
\AV\ ,
\eqn\suppd{\eqalign{W_{D5\; branes} & = \int^{u}_{u^*} log\frac{1-
e^{u-t}}{1-e^{-u}}du +  \int^{-u-t}_{u^*}
log\frac{1- e^{u-t}}{1-e^{-u}}du\cr
&= - 2 \sum_{n=1}\{\frac{e^{- n u}}{n^2}+\frac{e^{- n (t-u)}}{n^2}\}}}
The two contributions correspond to a single primitive disk
ending on the D6 brane and the image of this under $\Omega$ (see figure 2).

\bigskip
\centerline{\epsfxsize 4.0truein\epsfbox{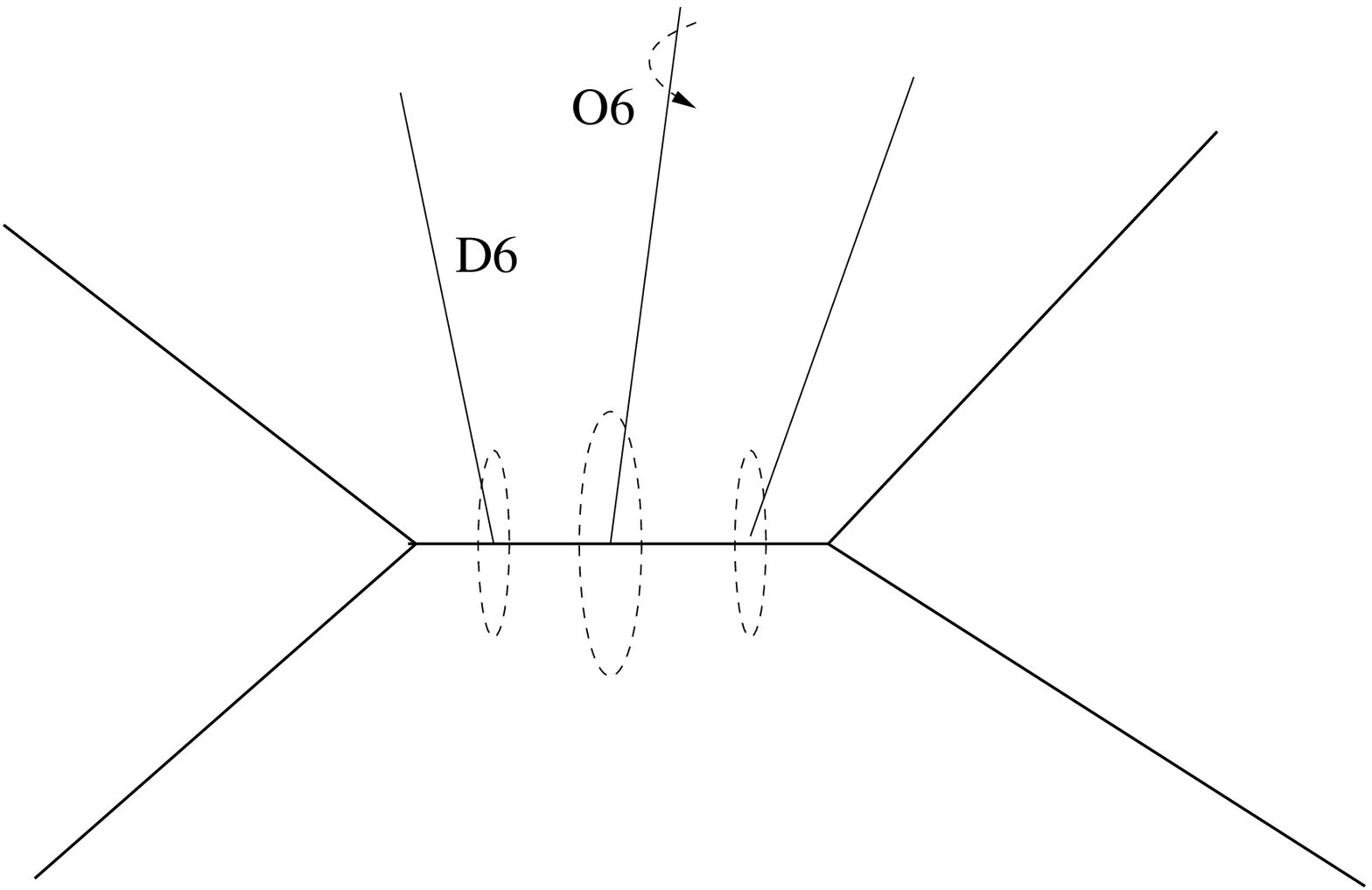}}
\rightskip 2pc \noindent{\ninepoint\sl
\baselineskip=8pt {\bf
Fig.2}: The orientifold of IIA on $O(-1)+O(-1) \rightarrow {\bf P^1}$.
The orientifold 6-plane has topology $C \times S^1$, and projects to a line
in the toric base. Superpotential receives contribution
from unoriented maps to the ${\bf CP^1}$.
In presence of additional D6 brane and its image, superpotential receives
contributions from the disk as well.}
\bigskip

Note that the result for $I^-$ has been obtained previously by
a different method \Sinha.  Namely, this orientifold of type IIA
string theory has been considered previously in the
context of duality with large $N$ $SO$ and $Sp$ Chern Simons theory on $S^3$
\Sinha, that generalized the original large $N$ conjecture
of \gv\ for unitary groups.
In this case, the free-energy of the Chern-Simons theory
\eqn\cs{F_{SO,Sp}(g_s,t) =
\pm\frac{1}{2} \sum_{n=1}^{\infty}
[1-(-1)^n] \frac{e^{-nt/2}}{2nsin(n g_s/2)} +
\frac{1}{2}\sum_{n=1}^{\infty} \frac{e^{-nt}}{n[2sin(n g_s/2)]^2}}
provides prediction
for all genus amplitudes of the $O(-1)+O(-1)\rightarrow
\bf{P^1}$ orientifold.
The first term in the above expression was interpreted in \Sinha\
as corresponding to contributions of all
non-orientable Riemann surfaces with a single
crosscap, as only odd powers of $g_s$ appear. Note that $\chi_{\bf
RP^2}=-1$ and that three crosscaps can be traded for a handle with a single
crosscap. For example, the partition function $F_{g=0,c=1}$
of the ${\bf RP^2}$ diagram is the coefficient of $g_s^{-1}$
and is easily seen to agree with the expression we obtained above
using mirror symmetry, up to over-all normalization.
The second term, on the other hand, corresponds to oriented maps, and in the
superstring language corresponds to superpotential like terms
generated from ${\cal N}=2$ amplitudes by turning on RR flux, and is
the $U(N)$ Chern-Simons amplitude, up to normalization, as discussed in
\Sinha .  It would be interesting to generalize the mirror symmetry
methods to also derive these higher genus predictions of large
$N$ duality for Chern-Simons theory.
\vskip 0.5cm

\newsec{Other Examples}
\vskip 0.5cm
It is easy to give more examples along
the lines we have discussed, which
involve more intricate contributions to the
superpotential.
Orientifold seven-planes have vanishing superpotentials,
and by mirror symmetry the disk amplitude of
D6 branes that are two dimensional in the
toric base vanishes as well.
To have a non-vanishing superpotential, the Type IIB theory must
have D5 brane charge, although this is only a necessary condition.
Thus we consider here examples whose mirror on the type IIB side
involve the orientifold operation $\omega \rightarrow \omega$.
Moreover, analogous to what was found in \refs{\AV,\AKV},
it is only the ``half''-orientifolds -- those whose fixed point set has
topology of $R^2$ in the B-model, that give non-zero superpotential in type
IIB string theory.

Since the setup is very similar to what we have done above,
we limit our presentation to showing on the figures in two
examples how the orientifold operation acts on the toric base.
In figure 3, the example depicts a case where there
is no superpotential generated, and figure 4 depicts a case
where there is a superpotential generated (as there is ``half''-orientifold
planes).

\bigskip
\centerline{\epsfxsize 3.5truein\epsfbox{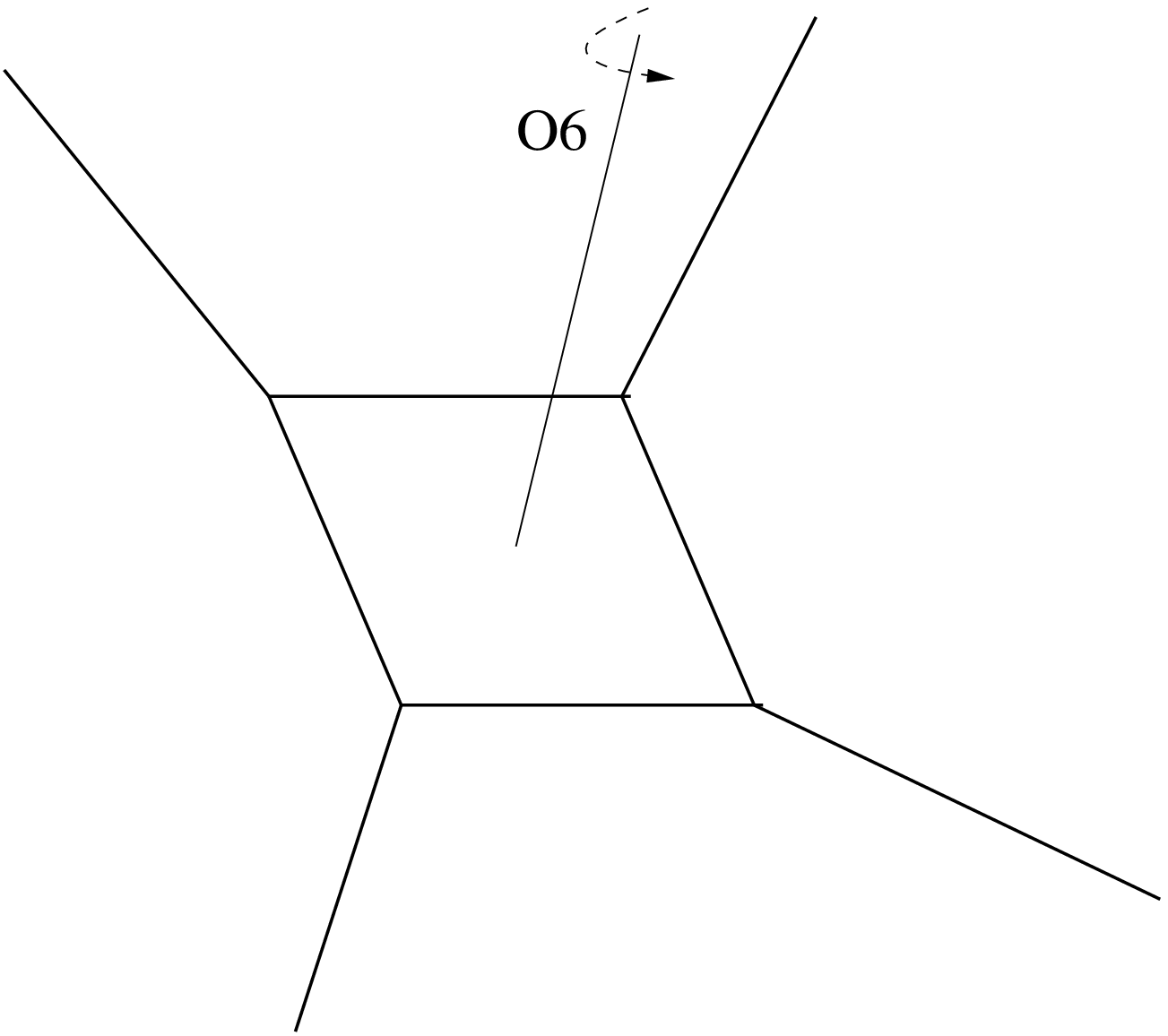}}
\rightskip 2pc \noindent{\ninepoint\sl
\baselineskip=8pt {\bf
Fig.3}: The orientifold of IIA on $O(K) \rightarrow {\bf P^1}\times {\bf
P^1}$ that acts as $I\Omega :\;\;(|z_1|,|z_2|)\rightarrow(1/|z_1|,1/|z_2|)$.
The orientifold does not generate
superpotential and this is related to the fact the orientifold action
projected to the base, fixes a
line meeting the boundary at a point in the interior.}
\bigskip

\bigskip
\centerline{\epsfxsize 3.5truein\epsfbox{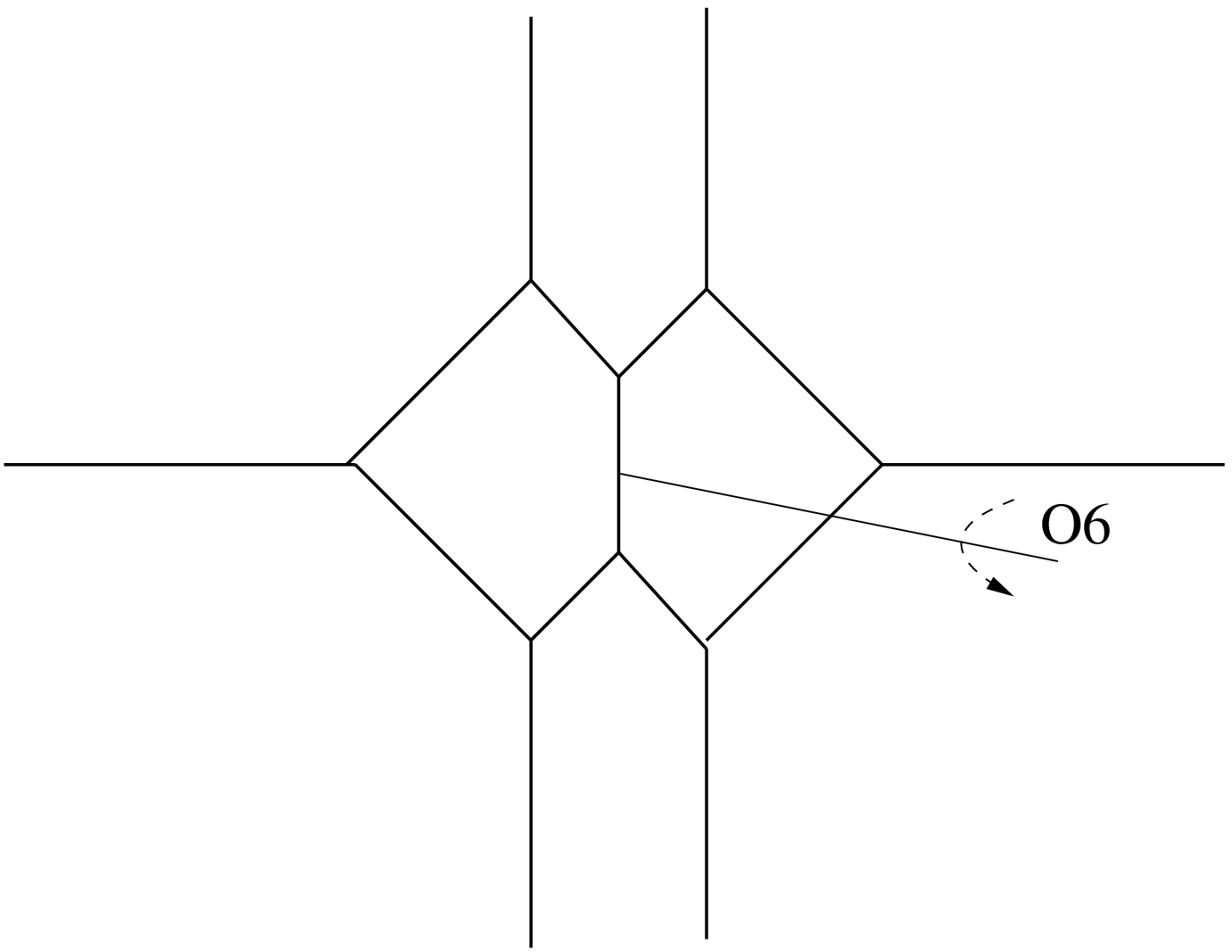}}
\rightskip 2pc \noindent{\ninepoint\sl
\baselineskip=8pt {\bf
Fig.4}: The orientifold of IIA that does generate superpotential.
This has $G=U(1)^5$. In addition to the superpotential,
$\bf{Z_2}$ action projects out $2$ out of five complexified Kahler
moduli.  }
\bigskip
\lref\SW{N.~Seiberg and E.~Witten,
``Gauge dynamics and compactification to three dimensions,''
arXiv:hep-th/9607163.}

\lref\DDHKMMS{J.~de Boer, R.~Dijkgraaf, K.~Hori,
A.~Keurentjes, J.~Morgan, D.~R.~Morrison and S.~Sethi,
``Triples, fluxes, and strings,''
arXiv:hep-th/0103170.}

It goes without saying that all these constructions
corresponding to type IIA orientifolds can also be lifted
up to M-theory involving $G_2$ holonomy metrics,
possibly with singularities in geometry.
It is known that
the simple $O6^-$ lifts up to smooth Atiyah-Hitchin manifold \SW~
 while $O6^+$ lifts to a D4-singularity that is frozen
by a discrete flux \refs{\Wtoro,\DDHKMMS}.
D6-branes may add more singularity of A and D-types.
In this context
both the disc instanton and the ${\bf RP^2}$ instanton correspond
to oriented Eulidean M2 brane instantons.

\centerline{\bf Acknowledgements}
We would like to thank I.~Brunner and R.~Myers for discussions.
B.A. is grateful to Harvard University, and 
M.A. and C.V. to Rutgers University where some of this
work was done. The research of M.A., K.H. and C.V.
is supported in part by NSF grants PHY-9802709 and DMS 0074329.
\listrefs
\end